\documentclass[journal]{IEEEtran}

\usepackage{graphics}
\usepackage{graphicx}
\usepackage{amsmath, amsthm, amssymb}
\usepackage{subfigure}

\newcommand{\E}{\mathbb{E}}

\newtheorem{definition}{Definition}

\newtheorem{counterexample}{Counter Example}
\newtheorem{thm}{Theorem}

\theoremstyle{remark}

\makeatletter
\@addtoreset{rem}{theorem}
\makeatother

\begin{document}

\title{\LARGE \bf Fundamentals of Large Sensor Networks: Connectivity, Capacity, Clocks and Computation}

\author{Nikolaos M. Freris, Hemant Kowshik, and P. R. Kumar\thanks{Nikolaos M. Freris, Hemant  Kowshik,  and P. R. Kumar are with the Dept. of ECE and CSL, University of Illinois at Urbana-Champaign, 1308 W. Main Street, Urbana, IL 61801, USA. Email:
{\tt nfreris2@illinois.edu}, {\tt kowshik2@illinois.edu}, {\tt prkumar@illinois.edu}.}
\thanks{This material is based upon work partially supported by by AFOSR under Contract No. FA9550-09-0121, USARO under Contract Nos. W911NF-08-1-0238 and W-911-NF-0710287, and NSF under Contract Nos. ECCS-0701604, CNS-07-21992, CNS-0626584, and CNS-05-19535.}}

\maketitle

\abstract
Sensor networks potentially feature large numbers of nodes. The nodes can monitor and sense their environment over time, communicate with each other over a wireless network, and process information that they exchange with each other. They differ from data networks in that the network as a whole may be designed for a specific application. 

We study the theoretical foundations of such large scale sensor networks. We address four fundamental organizational and operational issues related to large sensor networks - connectivity, capacity, clocks and function computation. 

To begin with, a sensor network must be connected so that information can indeed be exchanged between nodes. The connectivity graph of an ad-hoc network is modeled as a random graph and the critical range for asymptotic connectivity is determined, as well as the critical number of neighbors that a node needs to connect to. Next, given connectivity, we address the issue of how much data can be transported over the sensor network. We present fundamental bounds on capacity under several models, 
as well as architectural implications for how wireless communication should be organized.

Temporal information is important both for the applications of sensor networks as well as their operation. We present fundamental bounds on the synchronizability of clocks in networks, and also present and analyze algorithms for clock synchronization. 
Finally we turn to the issue of gathering relevant information, that sensor networks are designed to do. 
One needs to study optimal strategies for in-network aggregation of data, in order to reliably compute a composite function of sensor measurements, as well as the complexity of doing so. We 
address the issue of how such computation can be performed efficiently in a sensor network and the algorithms for doing so, for some classes of functions.

\normalfont

\section{Introduction}
Wireless sensor networks are composed of nodes with sensing, wireless communication, and computation capabilities. They can potentially deploy large numbers of sensors. The sensors themselves can measure the environment over time and generate data, from which we seek to extract relevant information. The sensor nodes communicate with their neighbors wirelessly, and cooperate with each other in processing the data. 

Sensor networks therefore feature a combination of many functionalities -- sensing, communication and computation, which are potentially exercised over large numbers of nodes. In this paper, we study the fundamental properties of such large scale sensor networks.

Since sensor networks assess the environment based on interacting with each other, a fundamental global property of the network  that is of interest is connectivity, which ensures that all nodes can communicate with each other over multiple hops. We study how it can be achieved through local properties of nodes. Modeling the  locations of sensor nodes as randomly distributed over a domain, we study how a connected sensor network can result from choices of communication ranges or by choices of neighborhood sizes, made locally by nodes.

Nodes in large scale sensor networks communicate with each other wirelessly. It is of interest to understand how the information that can be communicated scales with the number of nodes. We present different models for doing so, at different modeling granularities. We also address the issue of the architecture of the wireless network that can facilitate data transfer.

In sensor networks, the goal is often not to download all the measurements of all the nodes, 
but only to obtain certain concise functions of the data. For this purpose, one can exploit the computational capabilities of the nodes to process information as it flows through them so as to efficiently deliver only what is sought by the application for which the sensor network has been designed. It is therefore of interest to determine how to efficiently conduct such in-network information processing. We formulate and analyze this issue as one of function computation over a wireless network and present results on how this can be done efficiently, and what rates of data aggregation are possible for a large network.

The phenomenon being monitored by a sensor network is often time-varying, and so it is important to time-stamp events accurately. In some problems, such as localization, the accuracy of time-stamps is critical to the accuracy of the inference. This requires clock synchronization over the network, which is also important to operating the network efficiently, for example in synchronizing wake and sleep cycles or in scheduling other events. More generally, sensor networks are cyberphysical systems, where the notion of time is important for the physical system. We address fundamental issues related to what can and cannot be estimated vis-a-vis time and delays. We also present algorithms for clock synchronization and delay estimation, and study their properties for large networks.

A more complete understanding of fundamental issues that arise in large scale sensor networks can potentially provide a strong theoretical foundation for sensor networks, that can inform network designers about optimal design and operation of this new technology.

In Section \ref{sec_connectivity}, we study the key property of connectedness. We study the connectivity of a randomly deployed ad-hoc network using geometric random graph models and find the critical range beyond which the graph is connected with high probability. We also characterize connectivity based on the number of neighbors of each node, which permits the use of local logical variables to ensure the global property of connectivity.

In Section \ref{sec_capacity}, we present sharp order bounds on the amount of information that a wireless network can transfer under two distinct interference models, namely the protocol and physical models. We characterize the \emph{transport capacity} measured in \emph{bit-meters/sec}, where we say that the network transfers one bit-meter/sec, if one bit of information is transferred one meter closer to its destination within one second. In the case of a random network, we define the \emph{throughput capacity} to measure the achievable throughput per node (in \emph{bits/sec}). We also show how network information theory can be used to study the capacity of planar networks, and establish fundamental connections between the physical propagation properties of the medium and the resulting capacity of the network. We show the critical role played by signal attenuation parameters, absorption constant and path loss exponent, and provide insights into order optimal architectures.

The next issue we address, in Section \ref{sec_clock_sync}, is time. 
We present different possible notions of synchronization in a network. An important role is played in clock synchronization by the fact that link delays are unknown in a wireless network. We study the synchronization of affine clocks, and present a fundamental impossibility result for the case of unknown asymmetric link delays, as well as characterize the uncertainty set in the parameter space. 
We present and analyze a fully decentralized scheme based on spatial smoothing of the estimated time differences between pairs of clocks, and analyze its convergence rate for different classes of graphs. 

In Section \ref{sec_computation}, we present the outline of a theory of in-network computation, where we study optimal strategies for in-network aggregation of data, in order to compute a function reliably. The goal is to enhance application performance by focusing on the joint operation of communication  and computing in a sensor network. We consider function computation in tree graphs, and present a zero-error block computation approach that is optimal. In Section, \ref{sec_comp_type}, we classify symmetric functions into type-threshold functions, exemplified by Max or Min, and type-sensitive functions, exemplified by Average. We present order-optimal strategies for computing both classes of functions in multihop networks. In Section \ref{sec_comp_bool}, we provide exact results for computing Boolean functions in collocated networks, using tools from information theory and communication complexity.  The most general information theoretic formulation of this problem presents formidable problems which have remained unsolved even for simple networks. We review existing results from information theory, and present some interesting results regarding computation over noisy channels.

\section{Connectivity of wireless networks}\label{sec_connectivity}
From the viewpoint of communications, sensor networks are wireless ad-hoc networks formed by nodes which communicate with neighboring nodes over a wireless channel. Nodes in the network cooperate in sending data concerning each other. An important desirable global property of such a wireless ad-hoc network is that it is \emph{connected}, i.e., it constitutes a connected graph. This is particularly important in sensor networks, where achieving a common application objective  may require communication among all the nodes. An important example is when some information concerning all the sensor nodes needs to be collected by a designated fusion node. 

One way to obtain a point-to-point communication link from a node $i$ to a node $j$, is to ensure that the ratio of the received signal from $i$ to the noise at node $j$ is greater than a certain threshold. Given the power at which node $i$ transmits and the attenuation properties of the medium as well as an upper bound on interference and noise, one can calculate if node $j$ can successfully receive node $i$'s transmission. This gives rise to an undirected \textit{connectivity graph} with the set of nodes as vertices, where an edge $(i,j)$ is present if node $j$ is reachable by node $i$. A node can increase its range and its set of neighbors by increasing its transmit-power level. 
For uniform operation and analysis of a large number of nodes, an abstraction is to suppose that all nodes pick a common transmit range, and consider the problem of determining the \textit{critical range} at which each node needs to transmit so as to guarantee asymptotic connectivity of the network.

Since ad-hoc networks have arbitrary topology, the corresponding connectivity graph is modeled as a random graph. There are two important random graph models.\\
\textbf{Erdos-Renyi Graphs} - $\mathcal{B}(n, p(n))$ is a probabilistic graph consisting of $n$ nodes, in which edges are chosen independently and with probability $p(n)$. The critical probability for asymptotic connectivity of Erdos-Renyi graphs, see \cite{Bollobas}, is given by the following:
\begin{thm}
If $p(n) = \frac{\log n + c(n)}{n}$, then the probability that $\mathcal{B}(n, p(n))$ is connected converges to one as $n \rightarrow \infty$ if and only if $c(n) \rightarrow +\infty$.
\end{thm}
In an Erdos-Renyi random graph, the event that there is an edge between $i$ and $k$ is independent of the event that there are edges between $i$ and $j$, and $j$ and $k$. However, this is not generally true in an ad-hoc network, since it does not capture the loss of connectivity in a network when nodes are far from each other. Therefore one turns to graphs where the presence of a link between two nodes is based on the distance between the nodes. Such a modeling approach can be traced back to Gilbert \cite{Gilbert} who may be regarded as a pioneer of continuum percolation theory \cite{MeesterRoy}. \\
\textbf{Random Geometric Graphs} - Let $\mathcal{D}$ be a disk in $\mathcal{R}^2$ having unit area. Let $\mathcal{G}(n, r(n))$ be the graph formed when $n$ nodes are placed uniformly and independently in $\mathcal{D}$, and two nodes $i$ and $j$ are connected by an edge if the distance between them is less than $r(n)$. Then the problem is to determine the range $r(n)$ of transmissions which guarantees the probability that $\mathcal{G}(n, r(n))$ is connected goes to one as $n \rightarrow \infty$. This problem is studied in \cite{GuptaKumar_conn} and \cite{Penrose}:
\begin{thm}
If $\pi r^2(n) = \frac{\log n + c(n)}{n}$, then the probability that $\mathcal{G}(n, r(n))$ is connected  converges to one as $n \rightarrow \infty$ if and only if $c(n) \rightarrow +\infty$.
\end{thm}
A thorough treatment of random geometric graphs can be found in \cite{Penrose}.

Connectivity is a global property of the network, and it is of interest to determine if there are other local properties, besides range, that can also be used to achieve connectivity in a wireless network.  One such quantity is the number of neighbors of a node, which has the desirable feature that it is a {\em logical} property that can be checked locally by each node. The following result was established in \cite{XueKumar}:
\begin{thm}
Let $\mathcal{G} (n, \phi_n)$ be the network formed when each node is connected to its $\phi_n$ {\em nearest neighbors}, i.e., there is an edge $(i,j)$ if either $i$ or $j$ is one of the $\phi_n$ nearest neighbors of the other. Then there exist constants $c_2 > c_1 >0$ such that $\mathcal{G} (n, \phi_n)$ is connected (resp. disconnected) with probability approaching one as $n \rightarrow +\infty$ if $\phi_n \geq c_2 \log n$ (resp. $\phi_n \leq c_1 \log n$).
\end{thm}
The constants are refined in \cite{BalisterBollobas}, where it is shown that $c_2$ can be chosen smaller than one, indicating that fewer neighbors are needed for connectivity based on number of neighbors than range.

\section{Capacity of Wireless Networks}\label{sec_capacity}

Unlike wireline networks, a wireless network is a shared medium where transmissions interfere with one another.
Two or more nodes can make concurrent successful transmissions provided there is no destructive interference at the receivers. Two popular interference models are the following. In the \emph{protocol model} every active transmission from a node $k$ to a node $l$ creates an interference zone  which is a disk of radius $(1+\Delta)\rho_{kl}$ centered at node $k$, where $\rho_{kl}$ is the distance between nodes $k$ and $l$, and $\Delta>0$.. Transmission from a node $i$ to some node $j$ is successful if it is not in the interference zone of any other active transmitter.  In the \emph{physical model}, a transmission from $i$ to $j$ is successful if the received signal-to-interference plus noise ratio (SINR) at $j$ is higher than a given threshold, i.e., if
\begin{equation}
SINR(j) := \frac{P_i\rho_{ij}^{-\alpha}}{N + \sum_{k \in \mathcal{T} \setminus \{i\} } \rho_{kj}^{-\alpha}} \ge \beta,
\end{equation}
where $P_i$ is the power level of the $i-$th transmission, $\mathcal{T}$ is the set of concurrently active transmitters, $N$ is the ambient noise power level, $\alpha > 2$ is the path loss exponent, and $\beta$ is the threshold for successful reception. Each transmitter $i$ needs to satisfy a power constraint $P_i \le P_{ind}$.

Consider a planar wireless network of $n$ nodes deployed in a disc of fixed area $A$, where each node can transmit at a throughput of $W$ bits/sec.

As a measure of performance we can consider the \emph{transport capacity} ($C_T$) of the network measured in \emph{bit-meters/sec}, where we say that a network transports one bit-meter of information, if one bit of information is transferred one meter \footnote{If we assume that the transport capacity is equitably divided, the per-node transport capacity is obtained by dividing the transport capacity by $n$.}.
An alternative manner of assessing the performance is by supposing that each node randomly picks a destination from among the other nodes, and then determining the \emph{throughput capacity}, which is defined as the largest common throughput that can be provided to each origin-destination pair, i.e., the max-min throughput.

In \cite{gupta} the dependence of the transport capacity  on $n$ was studied both for the case of arbitrary networks, as well as the asymptotic behavior as $n \to \infty$ of the throughput capacity for random networks in which $n$ nodes are independently and uniformly distributed in a two-dimensional disc of area $A$. A throughput capacity $\lambda(n)$ is said to be feasible in this latter case if there is a spatiotemporal scheme for scheduling transmissions such that each node can send an average of $\lambda(n)$ bits/sec to its destination in a multi-hop fashion, with probability approaching 1 as $n \to \infty$.

The main results under the two interference models are summarized below \cite{gupta}, \cite{agarwal}.
\begin{thm}
\begin{verbatim}

\end{verbatim}

\begin{enumerate}
  \item (Protocol model) For a network with optimal node placement as well as optimal choice of origin-destination pairs, $C_T = \Theta(W\sqrt{An})$ \footnote{We use the standard notation due to Knuth. We write $f(n) = O(g(n))$ if $\limsup_{n\to +\infty} \frac{|f(n)|}{|g(n)|} \le c_1$ for some $c_1 < \infty$, $f(n) = \Omega(g(n))$ if $\liminf _{n\to +\infty} \frac{|f(n)|}{|g(n)|} \ge c_2,$ for some $c_2 < \infty$, and $f(n) = \Theta(g(n))$ if both $f(n) = O(g(n))$ and $f(n) = \Omega(g(n))$.}. In fact, for each $n$, 
      $ \sqrt{\frac{1}{\pi}} \frac{W}{\sqrt{(1+\Delta)\sqrt{\Delta}\sqrt{2+\Delta}}}\sqrt{A}\sqrt{n}    \le C_T \le \sqrt{\frac{8}{\pi}} \frac{W}{\sqrt{(1+\Delta)\sqrt{\Delta}\sqrt{2+\Delta}}}\sqrt{A}\sqrt{n}. $
  \item (Physical model) For all networks, $C_T = O(W\sqrt{An})$. For an optimally designed network $C_T = \Theta(W\sqrt{An})$.
  \item (Protocol model) For a random network, $\lambda(n) = \frac{c_1W}{\sqrt{n\log n}}$ is feasible while $\lambda(n) = \frac{c_2W}{\sqrt{n\log n}}$ is not, for appropriate $0 < c_1 < c_2 < \infty$, both with probability approaching 1 as $n \to \infty$.
  \item (Physical model) For a random network, $\lambda(n) = \frac{c_1W}{\sqrt{n\log n}}$ is feasible while $\lambda(n) = \frac{c_2W}{\sqrt{n}}$ is not, for appropriate $c_1,c_2$ both with probability approaching 1 as $n \to \infty$.
\end{enumerate}
\end{thm}
The constructive proof of capacity \cite{gupta} shows that, in the protocol model, an optimal scheme is to group the nodes into small
cells and designate one specific node per cell to relay multi-hop packets traversing the cell. It is also nearly optimal even if all nodes use the same transmission range (which is chosen to be just high enough to guarantee network connectivity).

In \cite{agarwal} the transport capacity is shown to be $\Theta(\sqrt{n})$, even under a generalized physical model where each node uses adaptive coding to attain a bit rate of $B\log(1 + SINR(i))$. 
In \cite{franceschetti}, the feasible throughput for a random network is shown to be $\Theta(\frac{1}{\sqrt{n}})$ for the generalized physical model.

In order to assess what are the ultimate limits to how much information wireless networks can carry and how they should be operated, one needs to turn to network information theory. In \cite{xie} the capacity of planar networks was formulated and studied under a signal path loss attenuation characteristic of the form $e^{-\gamma \rho_{ij}}\rho_{ij}^{-\delta}$, where $\rho_{ij}$ is the distance, $\gamma \ge 0$ is the \emph{absorption coefficient} and $\delta > 0$ is the \emph{path loss exponent}. It should be noted that generally there is always some absorption in the medium, so $\gamma>0$ \cite{franceschetti2}. Thus, the received signal at node $j$ at time $t$ is
\begin{equation}
y_i(t) = \sum_{i\ne j}c_{ij}e^{-\gamma \rho_{ij}}x_i(t) + z_j(t),
\end{equation}
where $z_j(t)$ is White Gaussian Noise (WGN) with variance $\sigma^2$. Above, $\rho_{ij}$ is  assumed to be no less than some $\rho_{min} > 0$. The signal transmitted by node $i$ at time $t$ can depend on all causally acquired information, i.e.,  the messages it wants to send to other nodes, say $w_{ik} \in \mathcal{W}_{ik},$ for $k\ne i$, as well as its past receptions $y_i^t := \{y_i(s):s\le t\}$. This allows for general cooperation strategies among nodes, and accounts for the power of information theoretic results when they are obtainable.

For a network of $n$ nodes there are $n(n-1)$ possible source-destination pairs $(i,j), j\ne i$. Suppose that a rate $R_{ij}$ is supported for source-destination pair $(i,j)$. There are power constraints of either the form $P_i\le P_{ind}$, for each node $i$, or $\sum_{i=1}^n P_i \le P_{total}$. The transport capacity is defined as $\sup_{(R_{ij}, i\ne j : \mbox{ achievable rate vector})} \sum_{i=1}^{n(n-1)}R_{ij}\rho_{ij}$, where achievability of a rate vector is defined in an information-theoretic sense \cite{cover_thomas}. 

It was shown in \cite{xie} that for all planar networks with $\gamma>0$ or $\delta\ge3$, $C_T \le \frac{c_1(\gamma,\delta,\rho_{min})}{\sigma^2}P_{total}$,  From this it follows that
\begin{equation}\label{E}
C_T \le \frac{c_1(\gamma,\delta,\rho_{min})P_{ind}}{\sigma^2}n,
\end{equation}
which is essentially a $\Theta(\sqrt{An})-$law, since the area itself grows like $n$ in this case, whence $\Theta(\sqrt{An}) = \Theta(n)$.

Concerning the regime $\gamma = 0 $, it was established in \cite{xie} that for $\delta < \frac{3}{2}, C_T$ can be unbounded even for fixed $P_{total}$. Moreover, to illustrate possibilities, for $\frac{1}{2} < \delta < 1$, super-linear scaling of the transport capacity, in place of (\ref{E}), was shown to be possible. Both results are achieved by a strategy of coherent multi-stage relaying with interference subtraction, which effectively shows that nodes can cooperate over large distances using coherence and multi-user estimation when the attenuation is low. 

The regime where the scaling of the form (\ref{E}) holds was further extended in \cite{xie2}, where it was established that $C_T = \Theta(n)$ for $\delta>2$. For the case where $\gamma=0$ and $\delta \in [1,2]$, by using novel forms of cooperation between nodes, it was established in \cite{tse} that $\lambda(n) \ge cn^{1-\delta-\epsilon}$, for $\delta \in [1,\frac{3}{2}]$ and $\lambda(n) \le \frac{c'}{\sqrt{n}}$, for $\delta \in [\frac{3}{2},2]$, where $\lambda(n)$ is the per-node throughput for a random destination. These results suggest the potential of other forms of cooperation besides multi-hop relaying in this attenuation regime. The precise dependence of constants on $n$ is investigated in \cite{xie3}.

\section{Clock Synchronization}\label{sec_clock_sync}

Distributed clocks generally don't agree. Yet, clock synchronization requirements are required for several applications in sensor networks. Applications include coordinating events in a decentralized system \cite{convergence}, tracking, surveillance, target localization \cite{plarre}, data fusion, scheduled operations like power-efficient duty-cycling which is especially important for low power sensor nodes operation,, as well as in sensor-actuator methods where loops are closed over networks. Slotted communication protocols for random Medium Access Control (MAC) \cite{GS} as well as environmental monitoring applications also require accurate clock synchronization. More broadly, as we head towards the era of event-cum-time driven systems featuring the convergence of computation and communication with control, the need for well-synchronized clocks becomes increasingly important, affecting QoS, system performance and safety.

\subsection{Notions of network clock synchronization}

There are two possible goals of clock synchronization 
\cite{nfrer2},\cite{roberto_thesis}.

\begin{enumerate}
  \item \emph{Ordering of events}. The goal is to create a right chronology of the events in the entire network, where knowledge of the exact time instants is not required, yet an ordering of events that may occur at different nodes  has to be determined. This was thoroughly studied in \cite{lamport} where the notion of \emph{virtual clocks} was introduced.
  \item \emph{Synchronization}. The goal of synchronization is to estimate the relative time differences among a set of clocks in the network. This information can be then used to translate time-stamps from one clock to any other clock. It has the advantage that the translation mechanism does not create undesirable dependencies by resetting clocks in hosts \cite{scott}, and gives rise to the notion of \emph{relative clocks}. By selecting a particular node as a reference, one can even set all clock displays to agreement and attain a global definition of time.
\end{enumerate}

\subsection{Fundamental limits for synchronization of affine clocks}

Let us first consider the simplest model of \emph{affine} clocks \cite{nfrer1}. Denote the time display of a fixed \emph{reference
clock} by $t$, and assume that the display $\tau_j(t)$ of a clock $j$ at time $t$, satisfies
\begin{equation}\label{affine_model}
\tau_j(t) := a_jt + b_j.
\end{equation}
Here $a_j > 0 $, is called the \emph{skew}, and $b_j$ is the \emph{offset} of clock $j$ at the time 0 of the reference clock.

\subsubsection*{IV. B. 1. \  Model for packet delay} \label{delay_model}

Delays in packet delivery constitute a fundamental limitation in synchronizing clocks over wireless sensor networks since they can be much larger than the required synchronization precision. Suppose that a packet sent by node $i$ is received by node $j$ after a delay of $d_{ij}$ time units (measured in the time units of the reference clock, clock 1). The delays $\{d_{ij}\}$ are assumed to be unknown but fixed. In fact delay estimation is an important problem in its own right, as well as an intrinsic part of the problem of clock synchronization \cite{solis,arvind}.

By the word ``\emph{delay}'' here is meant the sum of all delays incurred by a packet after it is time-stamped by the transmitter and before the time-stamp is read by the receiver. This includes (cf. \cite{kopetz,FTSP,nfrer1}):

\begin{enumerate}
  \item \emph{Transmission delay}. This accounts for the processing time in the transmitter after time-stamping. 
  \item \emph{Propagation delay}. This can be estimated accurately say by GPS, or other position information, since it only involves the distance between the nodes.
  \item \emph{Receiving delay}. This accounts for the processing time in the receiver before time-stamping. 
\end{enumerate}

With the exclusion of the electromagnetic propagation delay, the other delays can depend on the communication and computation platforms of the nodes involved and the load experienced at them. Due to this heterogeneity, delays are generally not symmetric, i.e., $d_{ij} \ne d_{ji}$, or identical between links.

\subsubsection*{IV. B. 2. \ Pairwise clock synchronization} \label{pairwise}
\begin{figure}[thpb]
  \centering
  \includegraphics[totalheight=0.14\textheight]{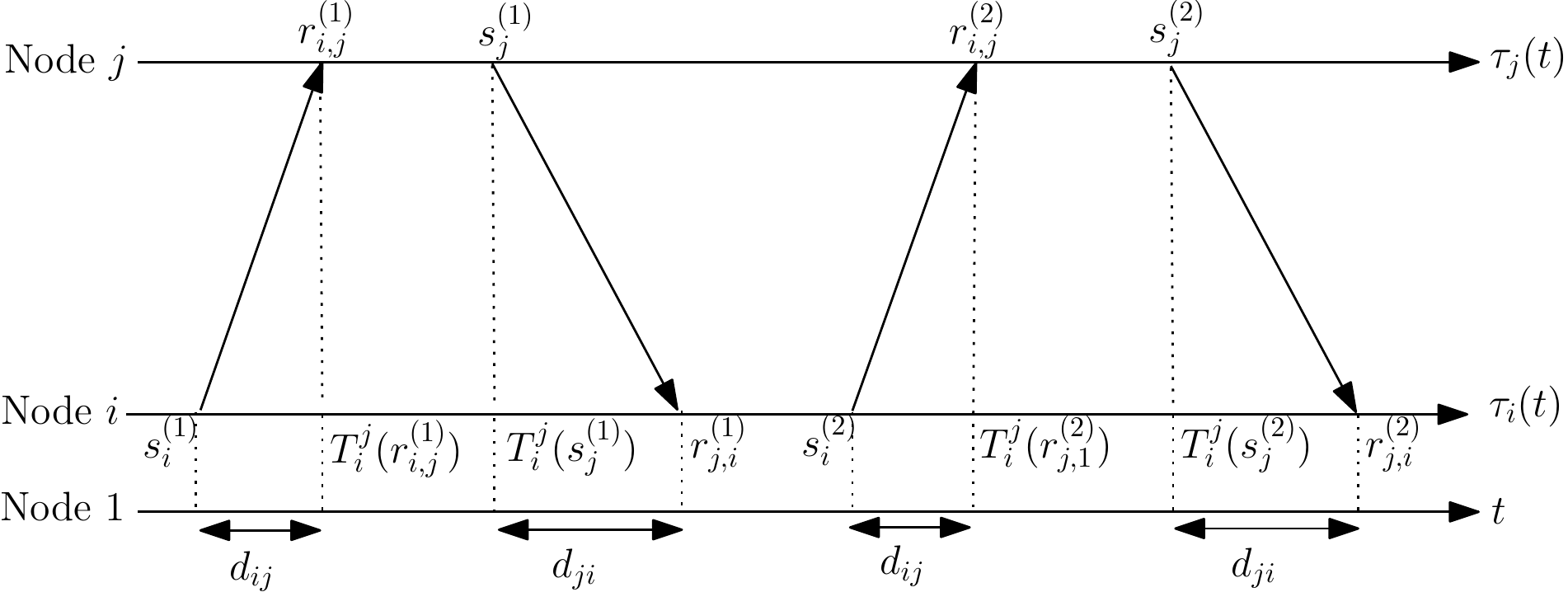}
  \caption{\small Message exchanges between two nodes}\label{ping1}
\end{figure}

Denote the time (as measured by the $i$-th clock) that node $i$ sends its $k$-th packet by  $s^{(k)}_i$ (see Figure \ref{ping1}), and denote by $r^{(k)}_{i,j}$ the time (as measured by the $j$-th clock) that node $j$ receives the $k$-th packet sent by node $i$.

It was shown in \cite{scott,nfrer1} that it is impossible to even synchronize just two clocks.
\begin{thm}
Even under bilateral exchange of an infinite number of packets between the two nodes $j$ and $1$, estimation of the entire four-tuple $(a_j, b_j, d_{1j},d_{j1})$ is \emph{impossible}. In particular,
\begin{enumerate}
  \item The skew can be estimated correctly, even if there are only two one-way pings, i.e., the link is
  unidirectional.
  \item The vector $(b_j,d_{1j},d_{j1})^T$ of offset and delays can only be determined up to a translate of a one-dimensional subspace of $\mathbb{R}^3$.
  \item The round-trip delay $(d_{1j}+d_{j1})$ can be estimated precisely.
  \item If we further use knowledge of {\em causality}, that packets cannot be received before they are
  sent, i.e., that $d_{ij}\ge 0$, and also that $a_j>0$, then the uncertainty set for the offset reduces to an interval,
  whose length is proportional to the round-trip delay.
\end{enumerate}
\end{thm}

If we assume that delays are symmetric in the two directions, the relative offset between the two clocks at the time of the receipt of the $k-$th packet from $j$ to $i, \tau_{ij}(k)$ (see Figure \ref{ping1}) can be estimated by
\begin{equation}\label{offset_est}
\hat{\tau}_{ij}(k) = -r^{(k)}_{ji} + (s^{(k)}_j + \hat{a}_{ij}\hat{d}_{ji}),
\end{equation}
where $\hat{d}_{ji}$ is the delay estimate given by
\begin{equation}\label{delay_est}
\hat{d}_{ij}(k) :=  \frac{1}{2}[(r^{(k)}_{ji} - s^{(k)}_j) + (r^{(k)}_{ij}-s^{(k)}_i) + (s^{(k)}_j - r^{(k)}_{ij})(1-\hat{a}_{ji})].
\end{equation}

\subsubsection*{IV. B. 3. \ Network clock synchronization} \label{pairwise}

Now consider the problem of synchronizing clocks over a network. Consider a network of $n$ nodes represented by a graph $G = (V,\E)$ which contains a directed edge from $(i,j)$ if $i$ can send packets to $j$. Node $1$ is considered to be the reference node. In the case of a link $(i,j)$ we call the ratio of the skews $\frac{a_j}{a_i}$ as the \emph{relative skew}, and the quantity $b_j - \frac{a_j}{a_i}b_i$ as the \emph{relative offset} between those two clocks. It is possible to determine the relative skew $\frac{a_j}{a_i}$ by use of two packets sent from node $i$ to node $j$, e.g., by $\frac{a_j}{a_i} = \frac{r_{i,j}^{(k+1)} - r_{i,j}^{(k+1)}}{s_i^{(k+1)} - s_i^{(k)}}$. The relative skew between two non-communicating nodes can be computed by multiplying the relative skews of the links across a directed path connecting those two nodes. In fact, every node in the network can estimate \emph{all} nodal skews if and only if the graph is strongly connected \cite{nfrer1}.

The following theorem \cite{nfrer1} establishes the result that without any further assumptions, clock synchronization is impossible in any network.

\begin{thm}\label{net_synch}
Consider a network of $n$ nodes. It is impossible to determine all
$2(n-1) + |\mathcal{E}|$ unknown parameters $\{a_i,b_i,$ and
$d_{ij}$ for all $i$ and all $j \ne i\}$ even if all pairs of nodes
can exchange any number of time-stamped packets containing any
information that is causally known to the transmitter. Furthermore, if the graph is strongly connected then:
\begin{enumerate}
\item All the skews $\{a_i:2\le i \le n\}$ can be estimated
correctly. \label{skew_correct}
\item Every vector $\underline{\hat{d}} = (\hat{d}_{ij}, (i,j) \in \mathcal{E})$
in the uncertainty set for the delay vector $\underline{d} =
(d_{ij}, (i,j) \in \mathcal{E})$ can be expressed as a known affine
transformation of $(n-1)$ variables $\{\hat{b}_i:2 \le i \le n\}$.
Each $\hat{b}_i$ can be regarded as an estimate of the unknowns
offset $b_i$. Any choice of these estimates $\{\hat{b}_i:2 \le i \le
n\}$ is consistent with all transmit and receipt time-stamps of all
packets.\label{offset_parametrization}
\item If causality is invoked, i.e., delays are non-negative, the uncertainty set for the
estimates of the offset parameters $\{\hat{b}_i:2 \le i \le n\}$ can
be fully characterized as a compact polyhedron of
$\mathbb{R}^{n-1}$. \label{polyhedron_characterization}
\item Suppose all links in $\mathcal{E}$ are bilateral. Then the feasible polyhedron in
\ref{polyhedron_characterization} has a non-empty interior if and
only if there is no bidirectional link with zero round-trip delay.
\label{nonempty_interior}
\end{enumerate}
\end{thm}


\subsection{Clock Synchronization algorithms}
We begin with the description of some well known clock synchronization protocols.
%
%
In large networks where synchronization requirements are not too stringent, e.g. Internet, the Network Time Protocol (NTP \cite{NTP}) has been used for over two decades. It is a hierarchical protocol with accuracy of the order of milliseconds \cite{NTP}, obtained by synchronizing with external sources organized in a hierarchy of levels, called \emph{strata}. Real-time applications in  wireless sensor networks typically require precision in the order of microseconds. In several cases, e.g., indoors, densely populated downtown areas, or during solar flares, the GPS service may be unavailable.

For more accurate synchronization in sensor networks and networked control a variety of algorithms have been suggested. Reference Broadcast Synchronization (RBS) \cite{RBS} is a receiver-receiver synchronization algorithm, which exploits the broadcast nature of the wireless medium. Nodes broadcast packets without any time-stamping on the transmitter side. The  nodes that receive the transmitted packets record the reception times and exchange them with their neighbors, so as to estimate their clock difference, by assuming that the one-way delays are the same for neighboring nodes. This completely eliminates the transmitter-side non-determinism, and accuracy depends mainly on the difference of receiving delays. This scheme was tested in actual sensor networks comprising of Berkeley motes and achieved precision within 11 $\mu s$ \cite{RBS}.

It was established in \cite{nfrer1} that all nodal skews are still determinable in the receiver-receiver synchronization scenario under some assumptions on the graph topology. However, the uncertainty set is a translation of $(2n-1)-$dimensional subspace and neither round-trip delays are determinable nor causality can be exploited (cf. Theorem 10, \cite{nfrer1}).

Timing-Sync Protocol for Sensor Networks (TPSN \cite{TPSN}) is a sender-receiver synchronization protocol which uses time-stamping at the MAC layer to eliminate the transmission delay which is typically the most variable term in wireless sensor networks. In \cite{TPSN}, it was observed and verified by simulations that TPSN achieves approximately twice the accuracy of RBS for pairwise synchronization. 
In \cite{FTSP}, authors identify further sources that contribute to packet delivery delay and propose the Flooding Time Synchronization Protocol (FTSP \cite{FTSP}). FTSP uses hardware solutions and efficient time-stamping to eliminate all packet delay factors but the propagation delay, and linear regression to compensate for clock drifts; the precision that it achieved was measured in the order of 10 $\mu s$ for a network with several hundreds of nodes. 

\subsubsection*{Network-wide offset estimation}

Given estimates of relative offsets for all links, i.e., between pairs of neighboring nodes, the network-level goal of synchronization is to obtain an estimate of all nodal offsets with respect to the reference node \cite{karp}, \cite{solis}, \cite{arvind}, \cite{barooah}.

Given relative offset measurements with some known variance,  the minimum variance unbiased linear estimate of the pairwise offset differences between any two nodes $(i,j)$ has been derived in \cite{karp}. It is shown that its variance is equal to the effective \emph{resistance} between those two nodes, in an electric network where each link's resistance equals the variance of the relative offset measurement. An iterative algorithm to obtain the minimum-variance nodal offset estimates is also derived.

A decentralized asynchronous algorithm based on \emph{spatial smoothing} of pairwise estimates has been developed and implemented in \cite{solis}, and comparative evaluations have been performed. The accuracy achieved was 2$\mu s$ for pairwise synchronization, and $20 \mu s$ for a network of $40$ nodes, in which FTSP achieved an average accuracy of $30 \mu s$. In \cite{arvind} the convergence of a synchronous version of this method was studied, and error asymptotics for various graph topologies were derived. A similar scheme which uses the Jacobi algorithm to obtain an iterative solution to a set of linear equations has been derived in \cite{barooah}.

We summarize the spatial smoothing algorithm of \cite{solis, arvind}.
In a network of $n+1$ nodes, where node $0$ is the reference node, denote the directed graph of links across which state estimates are available at some given time by $G = (V,\mathcal{E})$. Also denote the reduced incidence matrix of the graph obtained from the incidence matrix by removing the row corresponding to node $0$, by $A$ 
, and assume that the graph is connected. Then  $AA^T$ is the principal submatrix  of the \emph{Laplacian} of the graph, which is known to be positive definite for connected graphs, and hence invertible. Denote the relative offset on link $(i,j)$ by $\tau_{ij} := \tau_j - \tau_i$, where $\tau_i,\tau_j$ represent the local times of clocks $i,j$, respectively. We have
\begin{equation}\label{constraints}
o = A^Tv,
\end{equation}
where $o = \{\tau_{ij}\}\in \mathbb{R}^{|\mathcal{E}|}$ is the vector of relative offset estimates, $A\in\mathbb{R}^{|\mathcal{E}|\times n}$, $v = \{\tau_i\}_1^n \in \mathbb{R}^n$.
We can formulate the network-wide estimation problem as a least squares problem  with error criterion
\begin{equation}
F(v) := \|\hat{o} - A^Tv\|_2.
\end{equation}
The solution is $\hat{v}  = (AA^T)^{-1}A\hat{o} $. If we further assume that each link estimate is associated with a variance, then the nodal estimate $\hat{v}_i$ has variance equal to the effective resistance between node $i$ and node $0$, in an electric network where each undirected link has a resistance equal to the variance of its measurement. An application of the latter result, combined with circuit theory, allows us to determine the maximum error variance asymptotics for different graphs \cite{arvind}. In a tree network, the error variance is proportional to the diameter, while in a complete graph it decreases like $\Theta(1/n)$. In a lattice, the asymptotic growth rate is $\Theta(\log n)$.
What is very interesting, and  provides theoretical support for the feasibility of clock synchronization in planar wireless networks, is that the synchronization error can be kept bounded even in networks of large size \cite{arvind} 
\begin{thm}
Consider $n$ nodes located randomly (uniformly and independently) on a planar disk. Suppose that all nodes choose a common range that ensures connectivity (see section \ref{sec_connectivity}). Then, the error variance is bounded as the number of nodes $n \to \infty$.
\end{thm}

To avoid centralized computation of the pseudo-inverse operator $(AA^T)^{-1}A$, 
a distributed iterative implementation was derived in \cite{solis,arvind} as follows:
Setting the $i-th$ derivative of $F(v)$ equal to 0, yields $(AA^T)_iv - A_i\hat{o} = 0$, and a straightforward analysis \cite{arvind} shows that coordinate descent gives rise to an asynchronous scheme where node $i$ updates its estimate $v_i$ according to
\begin{equation}\label{SS}
v_i = \frac{1}{d_i} \sum_{j: (i,j)\in \mathcal{E} \mbox{ or } (j,i)\in \mathcal{E}} (v_j + \hat{o}_{ji}),
\end{equation}
where $d_i$, denotes the total degree of node $i$. Note that the scheme is fully distributed and has the further advantage that it uses no information about the network topology.

Above we have studied the asymptotic accuracy achievable in clock synchronization over various graphs. Next, we turn to the rate of convergence. In order to study convergence time of (\ref{SS}), \cite{arvind} considered a synchronous implementation of the form
\begin{equation}
\bar{v}_{k+1} = (I - D^{-1}AA^T)\bar{v}_k,
\end{equation}
where $D$ is the diagonal matrix of nodal degrees, and $\bar{v}_k := v_k - (AA^T)^{-1}A\hat{o}$. The convergence rate of the synchronous scheme is completely characterized by the spectral radius $\rho$ of $M:=(I - D^{-1}AA^T)$. Bounds on $\rho(M)$ were derived in \cite{arvind} by an application of Cheeger's inequality \cite{cheeger}:
\begin{thm}
The spectral radius $\rho(M)$ satisfies:
\begin{equation}
1 - 2\frac{d_0}{\sum_{i=1}^n d_i} \le \rho(M) \le 1 - (\frac{\kappa}{\sum_{i=1}^n d_i})^2,
\end{equation}
where $\kappa$ is the edge-connectivity of the graph. In particular, the settling time to an $\epsilon-$ neighborhood of the final value, for both lattice and random planar graphs, is $O(n^2)$.
\end{thm}

\section{A Theory of In-Network Computation}\label{sec_computation}
Sensor networks should be distinguished from general wireless ad-hoc networks. Traditional data networks are only concerned with end-to-end information transfer, but sensor networks may only be interested in gathering certain aggregate \emph{functions of distributed data}. For example, one might want to compute the average temperature for environmental monitoring, or the maximum temperature in fire alarm systems. In this context, communicating all the relevant data to a central collector node, which subsequently computes the function, might be inefficient, since it requires excessive data transfer or energy. Sensor nodes are severely limited in terms of power and bandwidth and it becomes necessary to find optimal \textit{in-network} aggregation and communication strategies for efficient function computation.  Such in-network processing and aggregation is made possible by the fact that sensor networks are often application specific, that is, deployed to achieve a specific goal, and hence nodes may look into the contents of packets and create new packets or discard others. This is in sharp contrast to data networks where nodes only process the headers of the packets, but never the contents of the payload of the packets.

A fundamental challenge therefore is to exploit the structure of the particular function of the data that is of interest, so as to optimally combine transmissions at intermediate nodes. Thus the problem of function computation is more complex than finding the capacity of a wireless network, since the traditional decode and forward model does not capture the possibility of combining information at intermediate nodes. The general problem of computing a function of correlated data, over a distributed network of nodes with wireless links, admits a variety of approaches. It is related to fundamental problems in multi-terminal information theory \cite{SlepianWolf} \cite{WynerZiv}, distributed source coding, communication complexity and distributed computation. In this section, we highlight the major approaches, and the hierarchy of ideas within.

\subsection{Block computation of functions in multi-party networks}\label{sec_comp_zero}
Consider a general network of $n$ nodes, with a designated collector node to which the value of the aggregate function of interest needs to be communicated. Each node has a certain transmission range, and can transmit directly to any other node within that range, provided no other transmission is interfering. Two networks of interest are the \textit{collocated} network, where the connectivity graph is complete; and the random \textit{multihop} network where the connectivity graph is a random geometric graph (see Section \ref{sec_connectivity}). This problem was studied in \cite{GiridharKumar}, to determine the limits of what can be achieved even after allowing for block computation. Nodes make a block of $N$ measurements, and the collector or fusion node seeks to compute the block of function values. Communicating a block of measurements has been shown by Shannon \cite{Shannon} to be critical to achieving reliability with high throughput. Analogously, the measure of efficiency in sensor networks is the \textit{computational throughput}, which is defined to be the minimum number of time units required per computation, over all schemes, and over all block lengths.

In sensor networks, we are often interested in \textit{symmetric} functions which only depend on the data of a sensor, not its identity. These are defined as functions whose values remain unchanged even if  the values of its arguments are permuted. They include statistical functions like mean, median, maximum/minimum, and others which are completely determined by the histogram of the set of node measurements. A key property is that the histogram of two disjoint sets can be combined to give the histogram of the union. This suggests a divide-and-conquer algorithm where a spanning tree is constructed, and partial histograms are propagated from children to parents up the tree towards the root where the collector is located. Further, one can achieve optimal spatial reuse of the wireless medium, by choosing the transmission range appropriately, and pipelining the transmissions (\cite{GiridharKumar}, Thm. 1).
\begin{thm}
For a connected network with $n$ nodes and maximum degree $O(\log n)$, the computational throughput is  $\Theta (\frac{1}{\log n})$. In particular, for the random network on a unit area square, if the common transmission range is chosen to be $\Theta(\sqrt{\frac{\log n}{n}})$, then one can achieve a computational throughput $\Theta (\frac{1}{\log n})$ with high probability.
\end{thm}
The lower bound $O(\frac{1}{\log n})$ follows from the fact that representing the histogram itself requires $\Theta (\log n)$ bits, and the collector node can receive only a bounded number of bits per slot. An important consequence of this result is that the natural method of \textit{tree aggregation} is order-optimal for symmetric function computation in random networks.

For block computation of a general function of correlated measurements, one needs to find a quantity analogous to the histogram, which can be composed across disjoint subsets of nodes. Towards this end, one can abstract out the medium access control problem associated with a wireless network, and view the network as a directed graph with edges representing essentially noiseless wired links between nodes. Thereby, we can focus on strategies for combining information at intermediate nodes, and optimal codes for transmissions on each edge.

We begin by considering the simple two node problem. Suppose nodes $v_X$ and $v_Y$ have measurements $x \in \mathcal{X}$ and $y \in \mathcal{Y}$, where the alphabets $\mathcal{X}$ and $\mathcal{Y}$ are finite sets. Node $v_X$ needs to optimally communicate its information to node $v_Y$ so that a function $f(x,y)$, which takes values in $\mathcal{D}$, can be computed at node $v_Y$ with zero error. One can consider both worst-case and, if a probability distribution over inputs $(x,y) \in \mathcal{X} \times \mathcal{Y}$ is specified, average case performance metrics. It is easy to see that any feasible encoder at $v_X$ must necessarily \textit{separate} $x_1$ and $x_2$ if there exists $y^* \in \mathcal{Y}$ such that $f(x_1, y^*) \neq f(x_2, y^*)$. In fact, one can show \cite{KowshikKumar} that the optimal encoder is one which greedily combines inputs in $\mathcal{X}$ that need not be disambiguated to obtain a reduced alphabet, and then assigns a distinct codeword to each input in the reduced alphabet. This specifies the optimal encoder in the worst-case scenario.

In the average case scenario, any feasible encoder at $v_X$ must separate $x_1$ and $x_2$, if there exists $y^* \in \mathcal{Y}$ such that $f(x_1, y^*) \neq f(x_2, y^*)$ and $p(x_1, y^*)p(x_2, y^*) > 0$. In the case where $p(x, y) > 0$ for all $x$ and $y$, the optimal encoder is again obtained by greedily combining inputs that need not be disambiguated and then applying the Huffman code on the reduced alphabet. If on the other hand, the probability distribution does have $0$s, finding the optimal encoder is NP-Hard, but the asymptotic per-instance complexity is a quantity called the graph entropy $H_{G}(X)$, as shown in \cite{AlonOrlitsky}.

Interestingly, this idea can be extended to obtain optimal in-network aggregation strategies on directed tree graphs \cite{KowshikKumar}. Since each edge is a cut-edge, one can find the disambiguation requirements for each edge, assuming all the nodes at either end of the edge are collaborating. Further, one can define encoders at each node, recursively and in a way that is consistent, and show that the above necessary condition is indeed sufficient.

Consider a directed tree graph with the collector as the root node. Let $T_i$ denote the subtree with node $i$ as root. The encoder at an intermediate node $i$ greedily combines input combinations of nodes in $T_i$ that need not be disambiguated, to obtain a reduced alphabet. In the worst-case scenario, the encoder at node $i$ assigns a fixed length codeword to each element in the reduced alphabet. In the average-case scenario, the encoder at node $i$ constructs a Huffman code on the reduced alphabet. The decoder at an intermediate node $i$ assigns a nominal input value to the nodes in its subtree $T_i$, based on the received transmissions \cite{KowshikKumar}.
\begin{thm}
For a directed tree graph, the above coding strategy achieves zero-error function computation, and simultaneously minimizes the number of bits transmitted on each edge, both in the worst-case scenario, and in the average case scenario when $p(x,y) > 0$.
\end{thm}

The extension to directed acyclic graphs presents some challenges. A key difference from the tree case is the presence of multiple paths to route the data, which present different opportunities to combine information at intermediate nodes. One can derive an outer bound to the rate region by finding the disambiguation requirements for each cut of the directed graph. This outer bound is not necessarily tight. A natural achievable strategy is to activate a subset of edges constituting a tree and then apply the optimal strategy for tree aggregation. The tree rate points are extreme points of the rate region. By time-sharing between different trees, one can achieve any rate point in the convex hull of the tree rate points. However, this does not match the outer bound for even simple examples .
\begin{figure}[thpb]
\subfigure[Counterexample $1$]
{\includegraphics[totalheight=0.15\textheight]{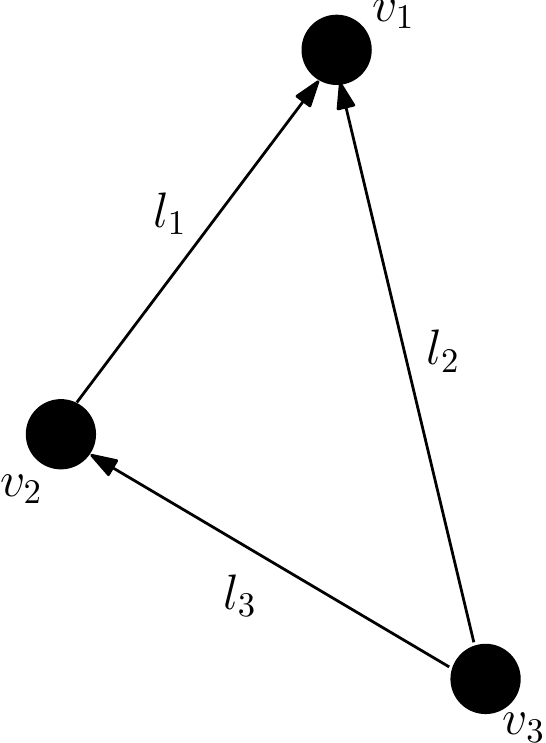}\label{simp_net}}
\hspace{.05in}
\subfigure[Parity in DAGs]
{\includegraphics[totalheight=0.15\textheight]{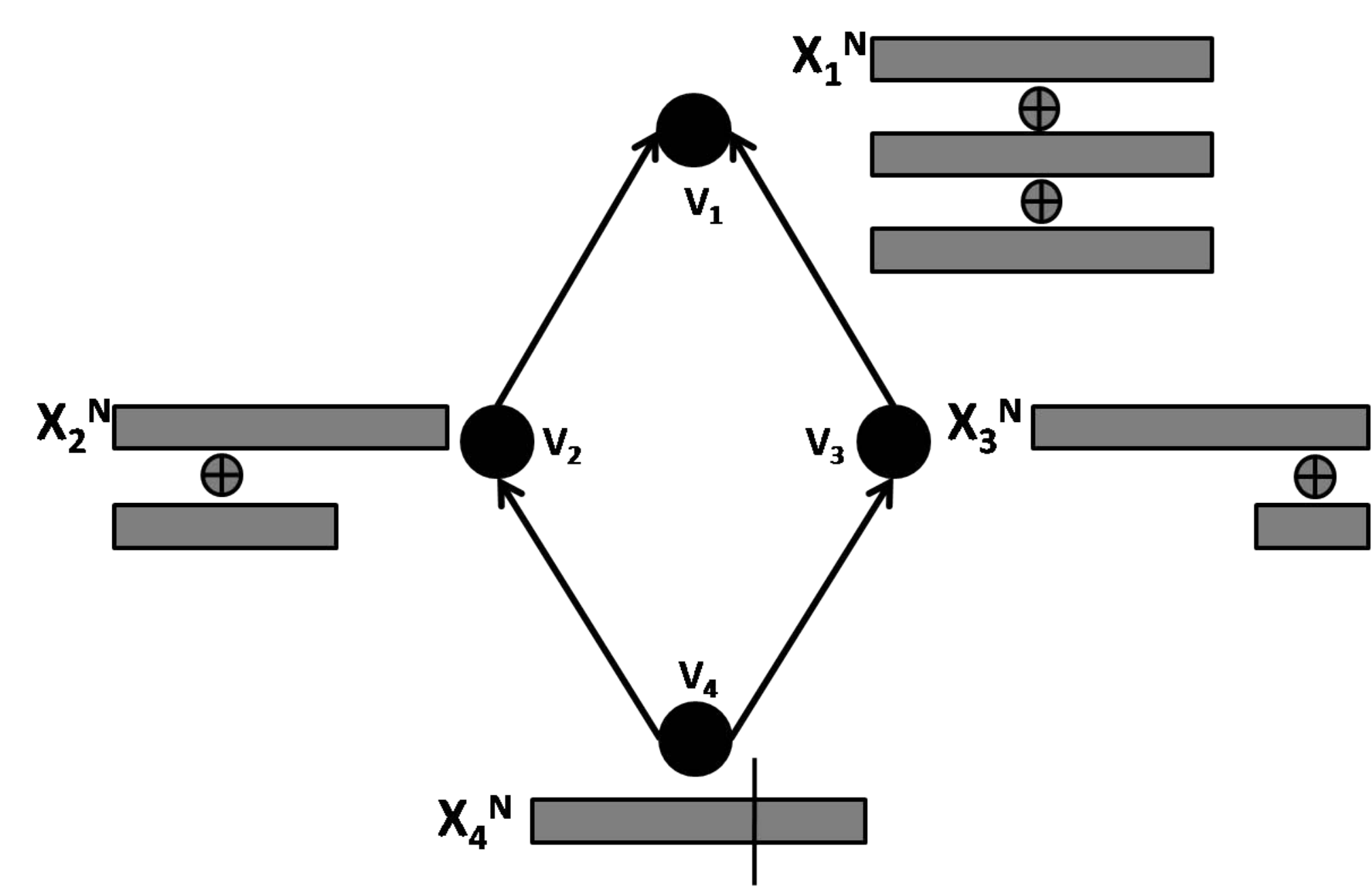}\label{simp_net2}}
\caption{Function computation in DAGs}
\end{figure}
\begin{counterexample}[Arithmetic Sum]\label{ex_arithmetic_sum}
Consider three nodes $v_1$, $v_2$, $v_3$ connected as in Figure \ref{simp_net}. Let $\mathcal{X}_2 = \mathcal{X}_3 = \{0,1\}$, with node $v_1$ having no measurements. Suppose node $v_1$ wants to compute $f(X_1, X_2, X_3) = X_2 + X_3$. Let $(R_{21}, R_{31}, R_{32})$ be the rate vector associated with edges $(l_1, l_2, l_3)$. The outer bound on $\mathcal{R}_{wc}$ is:\\
$\textrm{} \qquad  \qquad R_{21} \geq 1; \quad R_{21} + R_{31} \geq \log 3; \quad R_{32} + R_{31} \geq 1$.\\
The subset of the rate region achievable by trees is: \\
$R_{21} = \lambda + (1-\lambda)\log 3, R_{31} = \lambda, R_{32} = (1-\lambda)$ for $0 \leq \lambda \leq 1$.

Suppose that $X_1, X_2$ are i.i.d. with $p(X_1 = 0) = p(X_1=1) = 0.5$. The outer bound on $\mathcal{R}_{avg}$ is:\\
$\textrm{} \qquad  \qquad R_{21} \geq 1; \quad R_{21} + R_{31} \geq \frac{3}{2}; \quad R_{32} + R_{31} \geq 1$.\\
The subset of the rate region achievable by trees is: \\
$R_{21} = \lambda + (1-\lambda)\frac{3}{2}, R_{31} = \lambda, R_{32} = (1-\lambda)$ for $0 \leq \lambda \leq 1$.
\end{counterexample}

In some special cases, the cut-set outer bound can be achieved by aggregation along trees \cite{KowshikKumar}:
\begin{thm}
For the worst case computation of finite field parity in $\mathcal{F}_{D}$, and the maximum or minimum functions, the cut-set outer bound to the rate region is indeed tight.
\end{thm}
The idea is that every leaf node $v_i$ splits its block and transmits the segments on the outgoing edges from $v_i$ (See Figure \ref{simp_net2}). Each intermediate node receives partial blocks from lower nodes, and can hence compute an intermediate parity/intermediate maximum for some instances of the block. It then splits this intermediate processed block along the various outgoing edges. Proceeding recursively up the DAG, we can achieve the cut-set outer bound. 

The coding strategies described above can be viewed as a sophisticated form of network coding, which achieves efficient block computation. The multicast problem, where a source node wants to send the same message to multiple destinations, was studied in \cite{AhlswedeCai}. For this problem, linear network coding is known to achieve the cut capacity \cite{KoetterMedard}. The function computation problem is a somewhat dual problem, where the collector wants to calculate a function of sensor measurements, and we seek optimal encoders.

\subsection{Classification of symmetric functions}\label{sec_comp_type}
Among the class of symmetric functions, not all functions are equally hard to compute. For a function like Maximum, if a node knows that the maximum temperature recorded until now is at least $100$, then the node need not transmit any further reading unless it has a higher value. Thus the previous transmissions provide side information about the function value, even if the measurements are independent. On the other hand, for a function like Average, if even a single measurement is missing, we could have incorrect function computation, if the goal is zero-error block function computation. The results do change significantly if we allow a vanishing error as block length increases.

In \cite{GiridharKumar}, the collocated network scenario is analyzed, where all transmissions can be heard by all nodes, and collisions do not convey information. Thus, one restricts attention to \textit{collision-free strategies} where nodes can only transmit one at a time, with each node's transmission depending only on its data and previous transmissions all of which it has heard. Each node $i$ has a measurement $x_i$ taking values in a finite set $\mathcal{X}$. Define $\underline{x} := (x_1, x_2, \ldots, x_n)$, and let $\underline{\tau}(\underline{x})$ denote the \textit{type vector} of length $|\mathcal{X}|$. A symmetric function $f(\underline{x})$ depends on $\underline{x}$ only through its type vector $\underline{\tau}(\underline{x})$. Two interesting classes of symmetric functions can be identified, namely \textit{type-sensitive} and \textit{type-threshold} functions.
\begin{definition}[Type-sensitive function]
A symmetric function $f(\underline{x})$ is said to be type-sensitive if there exists some $\gamma$ with  $0 < \gamma < 1$, and an integer $\overline{n}$, such that for $n \geq \overline{n}$, and any $j \leq n - \lceil \gamma n \rceil$, given any subset $\{x_1, x_2, \ldots x_j\}$, there are two subsets of values $\{y_{j+1}, y_{j+2}, \ldots y_n\}$ and
$\{z_{j+1}, z_{j+2}, \ldots z_n\}$ such that
\begin{displaymath}
f(x_1, \ldots x_j, y_{j+1}, \ldots y_n) \neq f(x_1, \ldots x_j, z_{j+1}, \ldots z_n).
\end{displaymath}
\end{definition}
For a type-sensitive function, a certain minimum fraction of the arguments need to be known for the function value to be determined. Instances of type-sensitive functions include Average, Median, Majority and Histogram. Counting arguments can be used to show that the maximum rate for computing type-sensitive functions in the collocated network is $O(\frac{1}{n})$. A trivial achievable strategy of rate $\Omega(\frac{1}{n})$ involves each node declaring its value. This implies that the class of type-sensitive functions is orderwise maximally hard to compute.
\begin{definition}[Type-threshold function]
A symmetric function $f(\underline{x})$ is said to be type-threshold if there exists a non-negative threshold vector $\underline{\theta}$ of length $|\mathcal{X}|$, such that $f(\underline{x}) = f'(\underline{\tau}(\underline{x})) = f'(\min(\underline{\tau}(\underline{x}), \underline{\theta}))$, for all $\underline{x} \in \mathcal{X}^n$, with $min$ signifying element-wise minimum.
\end{definition}
Instances of type-threshold functions include Maximum, Minimum and $k^{th}$ largest value. In the case of binary measurements, for instance, the value of a type-threshold function is determined if the number of $0$s/$1$s exceeds a fixed threshold. Loosely speaking, the value of a type-threshold function is entirely determined by certain \textit{outstanding} measurements. For this class of functions, an exponential speedup is possible, 
and combinatorial arguments show that the computational throughput of this scheme is $\Omega(\frac{1}{\log n})$.

The upper bound on the computational throughput in each case is obtained by generalizing the concept of \textit{fooling sets} in communication complexity. Using this technique provides an upper bound on the throughput of all interactive protocols. We will revisit this idea in section \ref{sec_comp_bool}. Having derived an order-optimal strategy for collocated subnets, one can proceed to quantify the computational throughput of random planar networks. Spatial reuse of the wireless medium leads to a exponential speedup, and we have the following theorem \cite{GiridharKumar}.
\begin{thm}
The computational throughput of computing a type-sensitive function in a random planar network is $\Theta(\frac{1}{\log n})$; for a type-threshold function, it is $\Theta(\frac{1}{\log \log n})$.
\end{thm}

If we assume that the measurements are drawn independently and identically from some distribution, one can obtain an even higher computational throughput \cite{KowshikKumar}: 
\begin{thm}
Suppose that the measurements $\underline{X}_i = (X_{i,1}, X_{i,2}, \ldots, X_{i, N})$ are i.i.d. with $p(X_{i,l} = 1) = p$. Let $f(X_1, X_2, \ldots, X_n)$ be a symmetric type threshold function with threshold vector $[0, \theta]$. Thus the value of the function depends on the number of $1$s, upto a threshold $\theta$. The average case computational throughput for zero error block computation of the function $f$ is $\Theta(1)$ bits.
\end{thm}
\subsection{Computing Symmetric Boolean functions}\label{sec_comp_bool}
For certain special classes of functions, one can go further and seek exactly optimal strategies for computing functions in a collocated network. For the case of symmetric Boolean functions, this problem was studied in \cite{KowshikKumar_ITW}. 

Consider a collocated network with nodes $1$ through $n$, where each node wants to compute the function $f(X_1, X_2, \ldots, X_n)$ of the measurements. We seek to find communication schemes which achieve correct function computation at each node, with minimum worst-case total number of bits exchanged. Each node $i$ has a block of $N$ independent measurements, and we restrict ourselves to collision-free strategies. Let $\mathcal{S}_N$ be the class of collision-free strategies for block length $N$ which achieve zero-error block computation, and let $C(f, S_N, N)$ be the worst-case total number of bits exchanged under strategy $S_N \in \mathcal{S}_N$. The worst-case per-instance complexity of computing a function $f(X_1, X_2, \ldots, X_n)$ is defined by
\begin{displaymath}
C(f) = \lim_{N \rightarrow \infty}\min_{S_N \in \mathcal{S}_N} \frac{C(f, S_N, N)}{N}.
\end{displaymath}
We call this the \textit{broadcast computation complexity} of the function $f$.

Before we can address the general problem of computing symmetric Boolean functions, we consider the specific problem of computing the AND function, which is $1$ if all its arguments are $1$, and $0$ otherwise. The basic two node problem was studied in \cite{AhlswedeCai}:
\begin{thm}\label{thm_two_node_and}
Given any strategy $S_N$ for block computation of $X_1 \wedge X_2$,
\begin{displaymath}
C(X_1 \wedge X_2, S_N, N) \geq N\log_{2}3.
\end{displaymath}
Further, there exists a strategy $S_N^*$ which satisfies
\begin{displaymath}
C(X_1 \wedge X_2, S_N^*, N) \leq \lceil N \log_{2}3 \rceil .
\end{displaymath}
Thus, the complexity of computing $X_1 \wedge X_2$ is given by $C(X_1 \wedge X_2)=log_{2}3$.
\end{thm}
The lower bound is shown by constructing a \textit{fooling set} \cite{KushiNisan} of the appropriate size. Consider two nodes $X$ and $Y$, each of which take values in finite sets $\mathcal{X}$ and $\mathcal{Y}$, and both nodes want to compute some function $f(X,Y)$ with zero error. 
\begin{definition}[Fooling Set]
A set $E \subseteq \mathcal{X} \times \mathcal{Y}$ is said to be a fooling set, if for any two distinct elements $(x_1, y_1), (x_2, y_2)$ in $E$, we have either
\begin{itemize}
\item $f(x_1, y_1) \neq f(x_2, y_2)$, or
\item $f(x_1, y_1) = f(x_2, y_2)$, but either $f(x_1,y_2) \neq f(x_1, y_1)$ or $f(x_2,y_1) \neq f(x_1,y_1)$.
\end{itemize}
\end{definition}
Given a fooling set $E$ for a function $f(X_1, X_2)$, we have $C(f(X_1, X_2)) \geq \log_{2}|E|$. The extension to multi-dimensional fooling sets is straightforward and gives a lower bound on the communication complexity of the function $f(X_1, X_2, \ldots, X_n)$.

The above approach can be extended to the general AND function of $n$ variables, to obtain $C(\wedge(X_1, X_2, \ldots, X_n)) = \log_{2}(n+1)$. Further, one can derive the \textit{broadcast communication complexity} of a more general class of functions, called \textit{threshold functions}, which includes AND as a special case \cite{KowshikKumar_ITW}.\\
\begin{definition}[Boolean threshold functions]
A Boolean threshold function $\Pi_{\theta}(X_1, X_2, \ldots, X_n)$ is defined as
\begin{displaymath}
\Pi_{\theta}(X_1, X_2, \ldots, X_n) = \left\{\begin{array}{l} 1 \quad \textrm{if } \sum_{i}X_i \geq \theta, \\ 0 \quad \textrm{otherwise.}\end{array} \right.
\end{displaymath}
\end{definition}
\begin{thm}\label{thm_bool_threshold}
The complexity of computing a Boolean threshold function is $C(\Pi_{\theta}(X_1, X_2, \ldots X_n)) = \log_{2}\left(\begin{array}{c}n+1 \\ \theta \end{array}\right)$.
\end{thm}
We now turn to Boolean \textit{interval functions}, for which we only know the approximate complexity.
\begin{definition}[Boolean interval function]
A Boolean \textit{interval function} $\Pi_{[a,b]}(X_1, \ldots, X_n)$ is defined as:
\begin{displaymath}
\Pi_{[a,b]}(X_1, X_2, \ldots, X_n) = \left\{\begin{array}{l} 1 \quad \textrm{if } a \leq \sum_{i}X_i \leq b \\ 0 \quad \textrm{otherwise.}\end{array} \right.
\end{displaymath}
\end{definition}
A naive strategy to compute the function $\Pi_{[a,b]}(X_1, \ldots, X_n)$ is to compute the threshold functions $\Pi_{a}(X_1, \ldots, X_n)$ and $\Pi_{b+1}(X_1, X_2, \ldots, X_n)$. However, this strategy gives us more information than we seek, i.e., if $\sum_{i} X_i \in [a,b]^{C}$, then we also know if $\sum_{i} X_i < a$, which is superfluous information and perhaps costly to obtain. Alternately, we can derive a strategy which explicitly deals with intervals, as against thresholds. This strategy has significantly lower complexity \cite{KowshikKumar_ITW}. 
\begin{thm}\label{thm_bool_interval}
The complexity of computing a Boolean interval function $\Pi_{[a,b]}(X_1, X_2, \ldots, X_n)$ with $a + b \leq n$ is bounded as follows:
\begin{multline}
\log_{2}\left[\left(\begin{array}{c}n+1 \\ b+1 \end{array}\right) + \left(\begin{array}{c}n \\ a-1 \end{array}\right)\right] \leq C(\Pi_{[a,b]}(X_1, X_2, \ldots X_n)) \\
\leq \log_{2}\left[ \left(\begin{array}{c}n+1 \\ b+1 \end{array}\right) + (b-a + 1)\left(\begin{array}{c}n \\ a-1 \end{array}\right)\right]. 
\end{multline}
The complexity of computing a Boolean interval function $\Pi_{[a,b]}(X_1, \ldots, X_n)$ with $a +b \geq n$ is bounded as follows:
\begin{multline}
\log_{2}\left[\left(\begin{array}{c}n+1 \\ a \end{array}\right) + \left(\begin{array}{c}n \\ b+1 \end{array}\right)\right] \leq C(\Pi_{[a,b]}(X_1, X_2, \ldots X_n)) \\
\leq \log_{2}\left[ \left(\begin{array}{c}n+1 \\ a \end{array}\right) + (b-a+1)\left(\begin{array}{c}n\\ b+1 \end{array}\right)\right]. 
\end{multline}
\end{thm}

\subsection{Information theoretic formulation}\label{sec_comp_infotheory}
To study the ultimate performance limits of function computation and optimal function computation strategies, one again needs to turn to an information theoretic formulation. There are two features that can be incorporated in this most general formulation. First, we can allow for a vanishing error of computation, in contrast with the previous formulations which consider zero-error block computation. Second, we can exploit the correlation in sensor measurements, and achieve higher efficiency. However, there are very few results for this most general framework. We now review some of the basic information theoretic results, which lead to a more general formulation that incorporates computation over wireless networks.

Consider two sensors with measurements $X$ and $Y$ drawn according to the joint distribution $p(X,Y)$. The sensors take measurements in each time slot that are jointly correlated, but temporally independent and identically distributed. The two sensors are connected through noiseless independent links to a receiver, to which the measurements need to be communicated. The rates of the two links are $R_1$ and $R_2$ respectively. We wish to determine the \textit{rate region}, i.e., the set of all possible pairs of rates $(R_1, R_2)$ at which the sources can be individually compressed and sent to the receiver, such that the receiver can reconstruct the original sources with vanishing probability of error. This problem of \textit{decentralized compression of correlated sources} is known as the Slepian-Wolf problem \cite{SlepianWolf}.

The key challenge in this problem is to exploit the correlation in the measurements, even though the two sensors perform distributed compression. If both sensors had access to each others measurements, they would together still need to communicate at least $H(X,Y)$ bits, i.e, $R_1 + R_2 > H(X,Y)$. Further, if $X$ supposes that the receiver has full knowledge of $Y$, it still needs to transmit $H(X|Y)$ bits, i.e., $R_1 > H(X|Y)$. Applying a similar argument for $Y$, we have $R_2 > H(Y|X)$. The remarkable result of Slepian and Wolf is that this region is indeed achievable. The achievability strategy uses the technique of  \textit{random binning} to exploit correlation. The Slepian-Wolf problem can be easily extended to the case of multiple correlated sources communicating to a receiver. However, the extension to tree networks, with the receiver as root, has not been solved.


Another interesting variation of the problem arises if we only desire the reconstruction of sources to some \textit{fidelity}, i.e., the receiver wishes to recover estimates X, Y such that $E[D(X, X')] \leq d$, $E[D(Y,Y')] \leq d′$, where $D(\cdot, \cdot)$ is a given \textit{distortion measure}. This problem is open as well. The special case of this problem in which one of the sources is known to the receiver as \textit{side information}, and only the other is to be determined, was solved by Wyner and Ziv \cite{WynerZiv}. It is important to note that function computation can be viewed as a special case of the \textit{rate-distortion problem}, by defining an appropriate distortion metric that is function-dependent. For example, 
\begin{displaymath}
D(X, \hat{X}) := |f(X,Y) - \hat{f}(\hat{X}, Y)|^2.
\end{displaymath}

This has been extended in \cite{OrlitskyRoche} to the case in which the receiver desires to know a certain function $f(X,Y)$ of the single source $X$ and the side information $Y$, and determined the required capacity of the channel between the source and receiver as being a function of the conditional graph entropy, which is a measure defined on the two random variables and the characteristic graph \cite{Witsenhausen} defined by the function $f(\cdot)$. Recently, there have been some extensions to the case of two nodes \cite{DoshiShah} and to the case of tree networks \cite{FeiziMedard}. However, a general single-letter characterization has remained elusive.

Another interesting stand-alone problem was studied in \cite{KornerMarton}, where two correlated binary sources need to encode their sequences in distributed fashion, so that the receiver can compute the XOR function of the two sequences. It is shown that, in some cases, the sum rate required may be substantially less than the joint entropy of the two sources. This clearly displays the advantage of function-aware encoding strategies over standard distributed source coding.

The above formulation considers a single round protocol, where only $X$ communicates with $Y$. One can further seek the information theoretic limits for interactive computation. The rate region for multi-round interactive function computation has been characterized for two nodes in \cite{MaIshwar}, and for collocated networks in \cite{MaGuptaIshwar}. The characterization closely resembles the Wyner-Ziv result and some interesting connections with communication complexity are made.

\subsection{Modeling channel noise}\label{sec_comp_noisy}
All the above models assume that the channel is a noiseless wired link, possibly after the medium access control problem is solved. However, in practice, all channels are noisy, and one needs to study the rate of computation over noisy channels. In \cite{Gallager}, the problem of computing parity in a broadcast network was studied, assuming independent binary symmetric channels between each pair of nodes. In the noiseless case, one would require $n$ bits, and in the noisy case, by using repetition coding, we require $\Theta(n\log n)$ bits for low probability of error. However, this does not make use of the broadcast nature of wireless medium. For each bit that is transmitted, all the other nodes in the network hear a noisy version of it. The key contribution of \cite{Gallager} is a strategy which systematically uses this side-information to compute the parity function. The hierarchical strategy proposed requires only $\Theta(n\log \log n)$ bits. Remarkably, this upper bound of  $n \log \log n$ was shown to be sharp in \cite{Dutta}, using noisy binary decision trees.

In \cite{YingSrikantDullerud}, the problem of distributed symmetric function computation in noisy wireless networks is considered. It is shown that the energy usage for computing a symmetric function in a network with binary symmetric channels is $\Theta \left(n (\log \log n) \left(\sqrt\frac{\log n}{n}\right)^{\alpha}\right)$, where $\alpha$ is the path loss exponent.

While the information theoretic formulation described above only considers distributed source coding, one would like to generalize the formulation to incorporate a more general channel model. For example, consider two sources $S1$ and $S2$, which have access to channel inputs $X$ and $Y$ of a multiple access channel, with output $Z = X + Y + N$ being available to a receiver, where $N$ is Gaussian noise. The receiver wishes to compute the sum $X + Y$ . The question of interest is to find the optimal \textit{power-distortion curve}, i.e., for a given pair of transmit powers $P1$ and $P2$, what is the minimum distortion $D$ at which the sum $X + Y$ can be communicated to the receiver. Thus, the channel operation itself can be viewed as an implicit computation as argued in \cite{NazerGastpar}.

Finally, the above described solution to the medium access problem can be described broadly as \textit{interference avoidance}, since we assume that collisions do not convey information. We could generalize the formulation further by assuming an interference network. However, the solution to such a problem is very far from the current frontiers of knowledge in information theory.

\section{Conclusions}
We have presented some foundational results for sensor networks related to connectivity, capacity, clocks and computation. The fundamental approach studied is that of focusing on large networks and obtaining understanding through asymptotics. We have provided necessary and sufficient conditions that guarantee asymptotic connectivity in a wireless sensor network. We have presented the asymptotics on the capacity of wireless networks derived from both geometric and network information theoretic models and briefly discussed protocol implications. Next, we presented fundamental impossibility results on synchronizing the simplest class of affine clocks in networks, under idealistic assumptions, while fully characterizing the uncertainty set. We have presented a decentralized scheme for smoothing estimates of pairwise clock offsets. Finally, we have have addressed the broad problem of in-network function computation. We have presented optimal strategies for general functions on tree graphs, and studied tree aggregation in general graphs. We have also presented a classification of functions based on complexity of computation, and the exact complexity of computing certain Boolean functions in collocated networks. These results provide a basis for designing and analyzing large sensor networks.



\end{document}